\documentclass[doublecol]{epl2} 
\usepackage{amsmath}
\usepackage{bm}

\newcommand{\avg}[1]{\big\langle #1 \big\rangle}
\newcommand{\pd}{\partial}
\newcommand{\red}[1]{\textcolor{black}{#1}}
\newcommand{\blue}[1]{\textcolor{black}{#1}}

\title{Thinning and thickening in active microrheology}
\shorttitle{Thinning and thickening in active microrheology}

\author{Ting Wang\inst{1} \and Matthias Sperl \inst{1}}
\shortauthor{Ting Wang \etal}

\institute{                    
  \inst{1} Institut f\"ur Materialphysik im Weltraum,  
Deutsches Zentrum f\"ur Luft- und Raumfahrt (DLR), 51170 K\"oln, Germany\\
}


\abstract{
When pulling a probe particle in a  many-particle system
with fixed velocity, the probe's effective friction, defined as average pulling force over its
velocity, $\gamma_{eff}:=\langle F_{ex}\rangle/u$, first keeps constant (linear response), then decreases (thinning) and finally increases (thickening). 
We propose a three-time-scales picture (TTSP) to unify
thinning and thickening behaviour. 
The points of the TTSP are that there are three distinct time scales of bath particles: diffusion, damping, and  single probe-bath (P-B) collision;
the dominating time scales,
which are controlled by the pulling velocity,
 determine the  behaviour of the probe's friction.  We confirm the TTSP by Langevin dynamics simulation. Microscopically, we find that for computing 
the effective friction,  Maxwellian distribution of
bath particles' velocities works in low Reynolds number (Re) but fails in high Re. It can be understood based on the microscopic mechanism of thickening obtained in the $T=0$ limit. Based on the TTSP, we explain
different thinning and thickening observations in some earlier literature.
} 

\begin{document}

\maketitle

\section{Introduction}
Microrheology studies flow of complex fluids under micro-mechanical control \cite{Squires2010b,Puertas2014}.
It provides not only a novel method to understand materials' viscoelasticity  on the microscopic level
\cite{mason1995,Wilson2011,
Candelier2009a,Coulais2014}
but also a nice example of studying the response theory, which is a fundamental issue in statistical mechanics \cite{Kubo2003,evans2008,Marconi2008,Seifert2012}.
While in passive microrheology, only linear response to  thermal fluctuation is  possible, in active microrheology (AM), nonlinear response can also be realized by large pulling.        
A typical AM experiment is to pull a probe  particle
embedded in a complex fluid
with fixed velocity 
and then measure the fluctuating force  of it \footnote{One can also pull the probe with constant force, then measure the fluctuating velocity.}.
 The probe's friction coefficient, defined as average force  over its velocity $\gamma_{eff}:=\langle F_{ex}\rangle/u$, shows intriguing nonlinear behaviour: it first keeps constant (linear response) in the small velocity regime, then starts to decrease (thinning) in the moderate velocity regime, and finally may increase (thickening) in the large velocity regime. Similar behaviour can occur in bulk shear of macrorheology \cite{Seto2013,
wypart2014,
Kawasaki2014} as well.


Linear response and thinning 
were observed in colloidal systems 
both in experiments \cite{Wilson2009,Gomez-Solano2014}
and simulations \cite{Carpen2005,Winter2012}, theoretically studied by 
an effective two body Smoluchowski equation in low density  \cite{Squires2005b}, density functional theory \cite{Brader2014} and mode-coupling theory in high density
\cite{Gazuz2009,Gazuz2013,Gnann2011b}.  Thickening was observed in granular systems: both in static systems (bath particles at rest) in experiments
\cite{Takehara2010,Takehara2014} 
 and in driven systems (bath particles agitated by external driving) in simulation \cite{Fiege2012}, and was explained by a simple kinetic model \cite{Wang2014}.
However, a unifying description of thinning and thickening is still absent.
Why was only thinning observed in colloidal systems, while only thickening was observed in static granular systems? In this letter, we address this issue by proposing a TTSP.

\section{Three-Time-Scales Picture} 
\begin{figure}[t]
\begin{center}
  \includegraphics
  [width=0.9\linewidth]{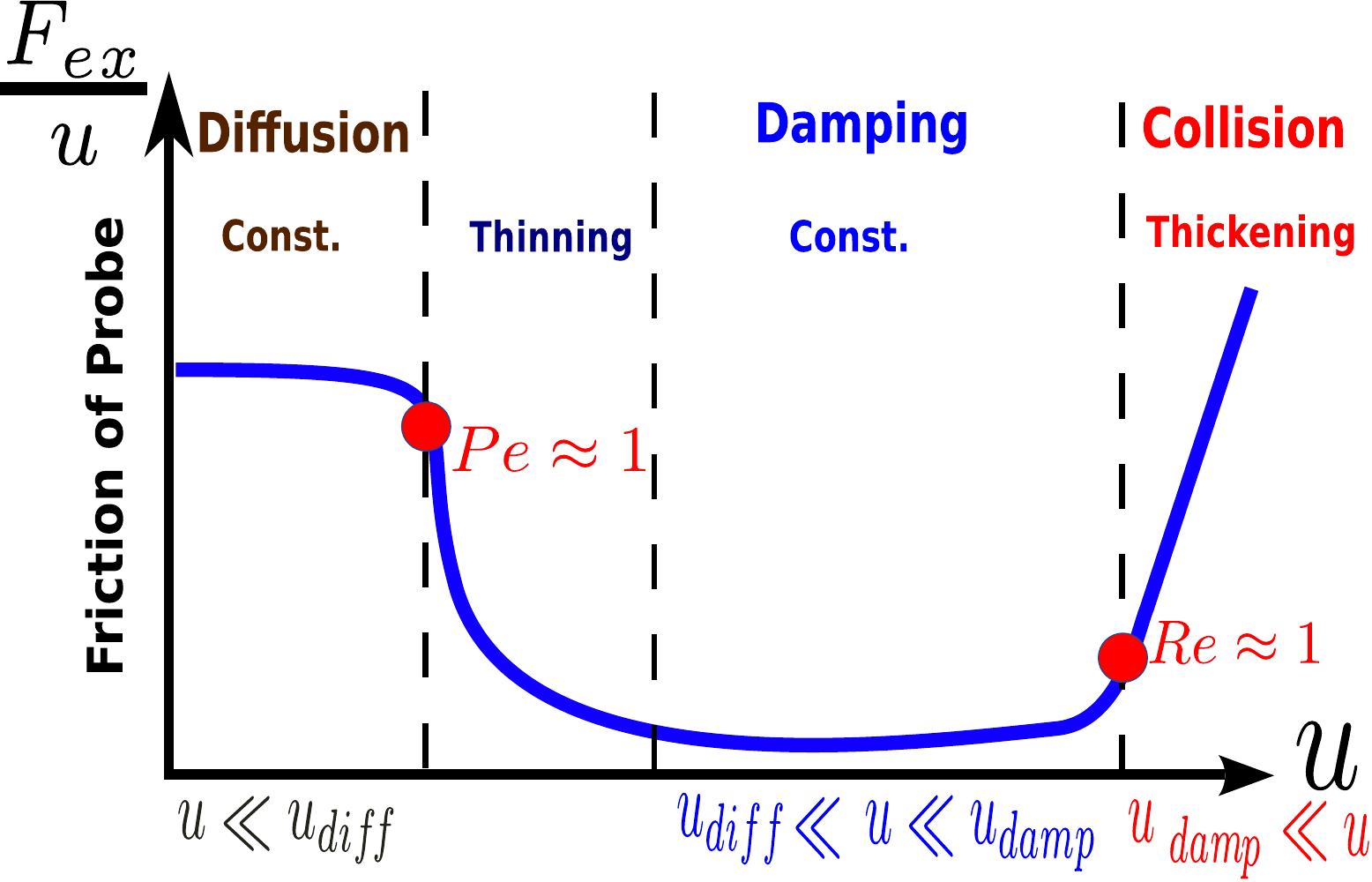}
\caption{\label{1}
Sketch of the TTSP of thinning and thickening: the effective friction  $\gamma_{eff}=F_{ex}/u$ vs. the    
pulling velocity $u$. 
Three time scales of bath particles are involved: diffusion, damping and single P-B collision, which lead to
three friction behaviour: a high plateau regime, a low  plateau regime, and an increasing friction regime, respectively.
The pulling velocity controls the dominating time scales. The crossovers cause thinning and thickening.
The turning points of thinning and thickening are around $Pe:=u/u_{diff}=\frac{R\gamma_0}{k_BT}u=1$ and $Re:=u/u_{damp}=\frac{m_b}{R\gamma_0}u=1$,   respectively.}
\end{center}
\end{figure}

A unifying picture of thinning and thickening 
is that 
three time scales of bath particles are involved (see fig.~\ref{1}.):
\begin{itemize}
\item  diffusion time scale:   $t_{diff}
=R^2/D$, where $D=\frac{k_BT}{\gamma_0}$ is the diffusion coefficient with the solvent friction $\gamma_0$, $R$ is the characteristic length scale (for hard sphere systems, it should be
the center distance of the probe-bath particles  contacting with each other). The corresponding diffusion velocity is $u_{diff}=R/t_{diff}$.

\item damping time scale:
$t_{damp}=m_b/\gamma_0$,  
where $m_b$ is the mass of a bath particle. The damping velocity is $u_{damp}=R/t_{damp}$.

\item  collision time scale: $t_{col}=R/u$, 
where $u$
is the pulling velocity. \blue{It characterizes
the mean-free time between first and second P-B collisions without damping.}  
The collision velocity is $u_{col}=R/t_{col}=u$.
\end{itemize}
The dominating time scales are controlled by the pulling velocity $\bm{u}$, 
which can be
 indicated by Peclet number $Pe:=u/u_{diff}=\frac{R\gamma_0}{k_BT}u$
and Reynolds number $Re:=u/u_{damp}=\frac{m_b}{R\gamma_0}u$. Different dominating time scales lead to different behaviour of the increased friction $\Delta \gamma_{eff}=\gamma_{eff}-\gamma_0$. In detail, (i) when the pulling velocity is small enough that $Pe\ll1$ and $Re\ll1$,
the diffusion dominates. $\Delta \gamma_{eff}$ arises from the diffusion of bath particles, which leads to a linear response regime.
 (ii) As the pulling velocity is much larger than the diffusion velocity but still much smaller than the damping velocity, \textit{i.e.} $Pe\gg1$ and $Re\ll1$,
diffusion is unimportant, damping dominates.
 $\Delta \gamma_{eff}$ arises from the damping of bath particles,
 which leads to another linear response regime. 
  (iii) As the pulling velocity is even larger than the damping velocity, \textit{i.e.} $Pe\gg1$ and $Re\gg1$, inertia dominates, $\Delta \gamma_{eff}$ arises from single P-B collision, which leads to an increasing friction regime.
  
The  plateau value of the linear response regime in (i) should be larger than the value in (ii), because diffusion causes larger friction in (i) comparing to the one arising from the damping only in (ii). As a result, the crossover from (i) to (ii) causes thinning. 
And the crossover from (ii) to (iii) causes thickening.
The turning points of thinning and thickening should be around $Pe=1$ and $Re=1$, respectively (see fig.~\ref{1}).

\section{Model}
To demonstrate the TTSP, we consider the model of
pulling  a probe particle with fixed velocity
embedded in 
a suspension of
N identical bath particles
\blue{in two dimensions.} Because  pulling with fixed force and pulling with fixed velocity 
behave similarly, both may show thinning and thickening 
\footnote{The effective friction of pulling with fixed velocity in general is larger than the one of pulling with fixed force  as pointed out in
\cite{Squires2005b} and further analysed in 
 \cite{Swan2013}},
we choose the latter for simplicity.
All particles are assumed to be
smooth and elastic hard disks with the same radius $r_0$. 
The dynamics of a bath particle (labelled $i$) and of the probe (labelled $p$)
obey the Langevin equations
\eqref{dyb} and \eqref{dyp}, respectively,
 \begin{subequations}
    \begin{align}
m_{b}\dot{\bm{v}}_i&=-\gamma_0\bm{v}_i
+\bm{\xi}_i+\bm{F}_{i,col} \label{dyb}
\\
0&=-\gamma_0\bm{u}
+\bm{\xi}_p+\bm{F}_{p,col}+\bm{F}_{ex}\label{dyp}
    \end{align}
  \end{subequations}
where 
$m_b$ is the mass of a bath particle;
$\bm{v}_i$ is the velocity of the $i$-th bath particle, $\bm{u}$ is the fixed pulling velocity of the probe;
$\gamma_0$ is the friction coefficient (all particles have the same value due to $\gamma_0\propto r_0\eta$; $\eta$ is the solvent's viscosity.);  
$\bm{\xi}_{k}$ ($k=i$ or $p$)
is a Gaussian random force 
satisfying the fluctuation-dissipation relation
$\avg{\xi_{k}^{\nu}(t)
\xi_{k'}^{\mu}(t')}=2\gamma_0 k_{B}T
\delta_{k,k'}
\delta^{\nu,\mu}\delta(t-t')$ $\,\big(\nu,\mu \in \{x,y\}$ are the components of the random force$\big)$; $\bm{F}_{i,col}$ (or $\bm{F}_{p,col}$)  is the interaction force between particles; $\bm{F}_{ex}$ is the external pulling force on the probe  only. 
\begin{figure*}[t]
\begin{center}

\includegraphics[scale=0.78]{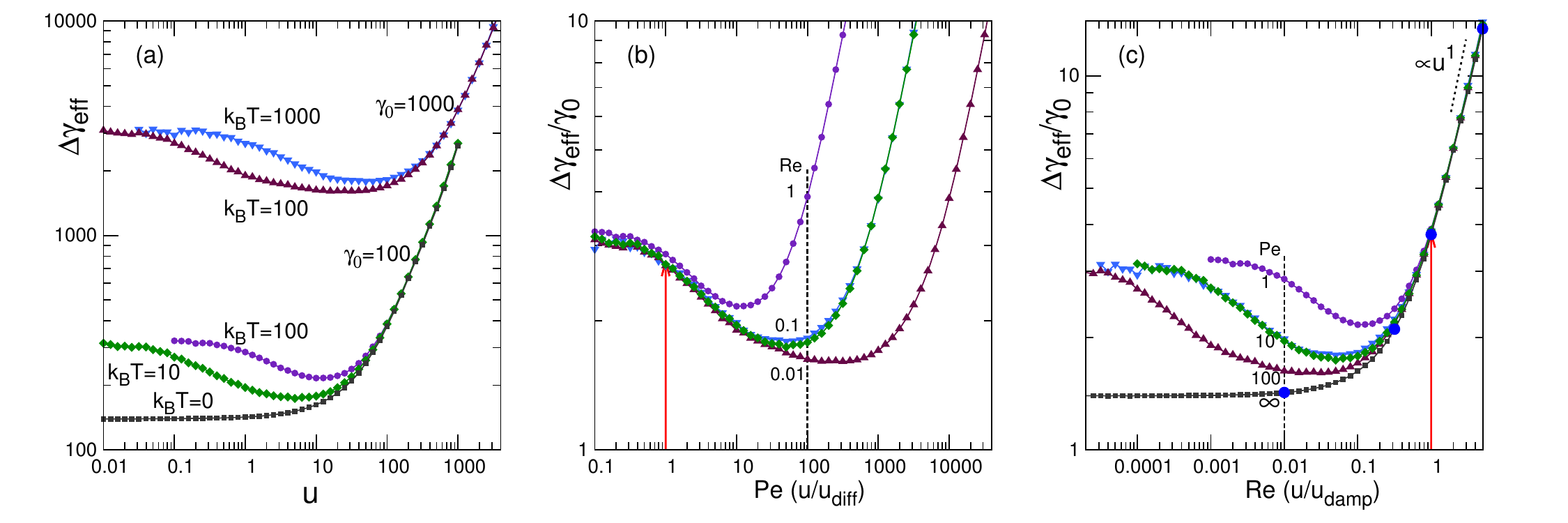}
\caption{\label{Pe_Re}
(a) Simulation results for the increased friction $\Delta\gamma_{eff}$ vs. the pulling velocity $u$ for different solvent frictions and temperatures.
\blue{
Different colors 
in (b) and (c) indicate different  solvent frictions and temperatures
as these  labelled in (a).$\,\,\,\,$} (b) 
The rescaled increased friction 
$\Delta\gamma_{eff}/\gamma_0$ vs. the
Peclet number  $Pe=u/u_{diff}=\frac{R\gamma_0}{k_BT}u$. The red arrow indicates the turning point of the thinning around  
$Pe=1$. 
\blue{At $Pe=100$, for 
different solvent frictions and temperatures, the corresponding value of the Re numbers are indicated.$\,\,\,\,$} 
(c) The rescaled increased friction 
$\Delta\gamma_{eff}/\gamma_0$ vs. the
Reynolds number  $Re=u/u_{damp}=\frac{m_b}{R\gamma_0}u$. The red arrow indicates the turning point of the thickening around  
$Re=1$.
\blue{At $Re=100$, for 
different solvent frictions and temperatures, the corresponding value of the Pe numbers are indicated. On the plot of $\{\gamma_0=100,\, k_BT=0\}$, four filled blue circles at $Re=0.01,\,0.3,\,1,\,5$
are drawn to 
compare with the corresponding streamlines in 
fig.~\ref{creep}.  
}
}
\end{center}
\end{figure*}
According to the equation of motion (EOM) \eqref{dyp}, 
the probe's increased friction  $\Delta\gamma_{eff}:=\gamma_{eff}
-\gamma_0$ is
\begin{equation}\label{gam} 
\Delta\gamma_{eff}
=\avg{F_{p,col}}/u,
\end{equation}
where $u,F_{ex}, F_{p,col}$ are absolute values of the corresponding vectors,
and the random force has been averaged out: $\avg{\bm{\eta}_p}=0$. 

Obviously, the P-B interaction directly leads to the increased friction,
while the bath-bath (B-B) particle interaction  affects  indirectly.
\emph{We omit
the B-B interaction in our model} because 1) 
such interaction may not be necessary for thinning and thickening behaviour; 2) the omission itself should be valid in the low density limit. 
In addition, we 
set the mass of the probe as much heavier than the mass of a 
bath particle: $m_p\gg m_b$, so that
in the coordinate of the probe, P-B collision just causes  specular reflection of the bath particles, but affects little the probe's velocity.  

Now the system of pulling a probe with fixed velocity $\bm{u}$ is equivalent to the system of a flow with velocity $-\bm{u}$ of a suspension of $N$ non-interacting bath particles passing a fixed disk with radius $R=2r_0$. 
The EOM of a bath particle (the index $i$ is dropped) 
in the coordinate of the probe is
\begin{subequations}\label{sto}
\begin{equation}\label{dyb2}
m_b\dot{\bm{v}}=
-\gamma_0 (\bm{v}+\bm{u})+\bm{\xi}
\end{equation} 
with the reflecting boundary condition (RBC)
\begin{equation}\label{rbc}
\bm{v}= \bm{v}-(\bm{v}\cdot\bm{e}_n)\bm{e}_n
\,\mbox{ for } |\bm{r}|=R,
\end{equation}
\end{subequations}
where 
$R=2r_0$ is the contact distance between the probe and a bath particle, and
$\bm{e}_n$ is the unit normal vector along   the direction from the center of the probe to that of the bath particle colliding with it.
Note that the P-B interaction term $\bm{F}_{i,col}$ in Eq.\eqref{dyb} is mapped into the RBC \eqref{rbc}.

Being equivalent to its 
stochastic description Eq.\eqref{sto},
the probability description of 
a bath particle obeys the 
 Fokker-Planck equation (FPE)
\begin{subequations}\label{fpe0}
\begin{equation}\label{fpe}
\pd_tP(\bm{r},\bm{v},t)=-\bm{v}\cdot\pd_{r}P
+\frac{\gamma_0}{m_b}\pd_{\bm{v}}\cdot
\Big[
(\bm{v}+\bm{u})+\frac{k_BT}{m_b}\pd_{\bm{v}}
\Big]P,
\end{equation} 
which can be obtained by Kramers-Moyal expansion
\cite{Risken1989} of Eq. \eqref{dyb2}. The corresponding  RBC is 
\begin{equation}\label{rbc2}
P(\bm{r},\bm{v},t)=P(\bm{r},\bm{v}-(\bm{v}\cdot\bm{e}_n)\bm{e}_n,t) \mbox{ for } |\bm{r}|=R.
\end{equation} 
\end{subequations}

In principle,  
the steady state equation ($\pd_tP=0$) of the FPE \eqref{fpe}  can be  solved with the 
RBC \eqref{rbc2}. Then one can obtain the  average collision force of $N$ bath particles on the probe:
\begin{equation}
\label{fcol}
 \begin{split}
 \avg{\bm{F}_{p,col}}=\int d\bm{v}\oint_{r=R}dl&
NP_{st}(\bm{r},\bm{v})\bm{v}\cdot (-\bm{e}_n)\Theta [\bm{v}\cdot (-\bm{e}_n)] 
\\
& 2 (-\bm{e}_n)m_b\bm{v}\cdot (-\bm{e}_n) 
 \end{split}
 \end{equation}
where $P_{st}$ denotes the steady distribution, $dlNP_{st}(\bm{r},\bm{v})\bm{v}\cdot (-\bm{e}_n)
\Theta [\bm{v}\cdot (-\bm{e}_n)] 
$ 
is the density current of bath particles with velocity $\bm{v}$ passing through a small contact surface
$dl\bm{e}_n$  ( $\Theta (x)=1$ for $x\geq0$;$\Theta (x)=0$ for $x\leq0$ ),  
and $ 2(-\bm{e}_n)m_b \bm{v}\cdot (-\bm{e}_n) $ is the bath particles' momentum transferred 
to the probe due to single P-B collision. Inserting 
Eq.\eqref{fcol} into Eq.\eqref{gam}, one obtains the effective friction $\gamma_{eff}$.

In practice, however, to analytically solve the FPE \eqref{fpe0} is difficult due to the RBC \eqref{rbc2}. 
Our strategy is to numerically solve Eq.\eqref{fpe0} by simulation of the  stochastic process Eq.\eqref{sto},
because of its equivalence to the FPE  \eqref{fpe0} and simplicity.

\section{Stochastic Simulation} 
To calculate the effective friction, the stochastic dynamics simulation according to Eq.\eqref{sto} is performed.  
The discrete form of the Gaussian random force is $\bm{\xi}=\sqrt{2\gamma_0 k_BT/h} (\xi_0^x,\xi_0^y)$, where 
$\xi_0^\mu$ ($\mu\in {x,y}$) is the standard Gaussian random number of the probability distribution function as
 $P(\xi_0^\mu)=\frac{1}{\sqrt{2\pi}}\exp(-\frac{
 \xi_0^{\mu\,2}}{2})$,
 and $h$ is the time step of the dynamics set to be $h=\frac{1}{2\gamma_0}$ for different solvent frictions.
The box size is set to be $Lx\times Ly=20R\times 20R $ with periodic boundary conditions, which is large enough to suppress finite size effects. 
 The mass of bath particles and the P-B contact distance 
are set to be unit values: $m_p=1$, $R=1$. The  density of bath particles  is also rescaled to unit value $n_0=1$, since it is not a control parameter in our model due to the assumption of non-interacting of bath particles. 
\red{The control parameters are the pulling velocity $u$, the solvent friction $\gamma_0 $ and the temperature $k_BT$,
which 
are applied to investigate the whole regime of different time scales.} 

Initially, bath particles are homogeneously distributed in space with Maxwellian distributed velocities. 
Then the probe is pulled along the x direction with total running time $10R/u$, which ensures that the bath particles around the probe reach the steady state.  
After a transient time, the steady average P-B collision force is computed 
by detecting the bath particles 
 passing through the boundary:
 $\avg{\bm{F}_{p,col}}=
 \frac{1}{\Delta t}\oint_{r=R} dl\int_{0}^{\Delta t} dt 2
 \big[
 m_p (-\bm{e}_n)
\bm{v}(t)\cdot (-\bm{e}_n)\big]\Theta [\bm{v}(t)\cdot (-\bm{e}_n)] 
 $, which is the simulation realization of the  collision force expressed in Eq. \eqref{fcol}.
The corresponding increased friction 
$\Delta \gamma_{eff}$ is obtained based on Eq. \eqref{gam}.

\section{Result}
Fig.~\ref{Pe_Re} (a) shows the  simulation result of the
increased friction $\Delta\gamma_{eff}$
versus the pulling velocity $u$
for different solvent frictions and
temperatures, 
   $\{\gamma_0=1000,\,k_BT=1000,100\}$
   and $\{\gamma_0=100,\,k_BT=100, 10,0 \}$.
All plots, except for $\{\gamma_0=100,\, k_BT=0\}$, exhibit linear response, thinning and thickening as expected by the TTSP. \red{For the exception,
only linear response
and thickening occur, because no diffusion but only damping and collision time scales are involved. 
}

\red{Fig.~\ref{Pe_Re} (b)
 \footnote{Data set of $\{\gamma_0=100,\, k_BT=0\}$ is not included in fig.\ref{Pe_Re} (b), because $u_{diff}=0$, no diffusion is involved.}
shows that
the rescaled increased friction
$\Delta\gamma_{eff}/\gamma_0$ versus
 Peclet number, $Pe=u/u_{diff}=\frac{R\gamma_0}{k_BT}u$.
In the small Pe regime $Pe<1$, the diffusion time scale dominates,
all plots coincide with each other in a plateau value.
With increasing  $Pe$, diffusion becomes less  important, all plots  start to
 decrease
around $Pe=1$,
which agrees with the  TTSP. 
\blue{Between $Pe=100$ and $Pe=1000$, for $\{\gamma_0=1000,\, k_BT=100\}$, the brown line, clearly there is a second plateau lower than the first one, being consistent with the TTSP.}
In addition, the length of the thinning regime varies for
different data sets
\footnote{\red{The exception is that $\{\gamma_0=1000,\, k_BT=1000\}$ and
$\{\gamma_0=100,\, k_BT=10\}$
 coincide with each other in  
 both fig.\ref{Pe_Re} (b) and (c),
 because for the same $u$, they have the same $Pe$ and $Re$ numbers.}},
because for the same $Pe$, the $Re$
numbers can also be different.
At $Pe=100$, 
for $\{\gamma_0=1000,\,k_BT=100\}$,
$Re=0.01$,
bath particles are still in the damping regime;
for  $\{\gamma_0=100,\,k_BT=100\}$, $Re=1$,
bath particles are already in the
inertia  (thickening) regime,
which suppresses the thinning process.
}

\red{Fig.~\ref{Pe_Re} (c)
shows that
the rescaled increased friction
$\Delta\gamma_{eff}/\gamma_0$ versus
 Reynolds number, $Re=u/u_{damp}=\frac{m_b}{R\gamma_0}u$. All plots start to converge around $Re=1$, which agrees with the TTSP. 
\blue{In the small $Re$ regime, for different plots, at $Re=0.01$, the frictions increase with the decreasing $Pe$ as indicated in the figure, which supports the TTSP that the diffusion causing larger friction than the one in the damping only regime $Pe\rightarrow\infty$.} 
 For $Re>1$, \textit{i.e.} the inertia regime, all plots coincide with each other and asymptotically tend to  $\Delta\gamma_{eff}
 \propto u$,  because the flux of bath particles passing through the P-B contact surface is $j\propto n_0u$ with momentum transferring to the probe  $p\propto m u$,
 and $\Delta\gamma_{eff}=\avg{F_{col}}/u=j p/u\propto u$.} 

\red{
In summary,  
the friction behaviour of different dominating time scales 
and of the two turning points as shown in  fig.~\ref{Pe_Re}, all agree quite well with the TTSP.
}







\section{Microscopic picture}
\blue{The TTSP is indicated by $Pe$ and $Re$.}
\red{
Microscopically,
what happens in the different $Pe$ and $Re$
regimes?}
\subsection{\blue{
density distribution}}
\begin{figure}[htbp]
\begin{center}
 \includegraphics[width=1\linewidth]{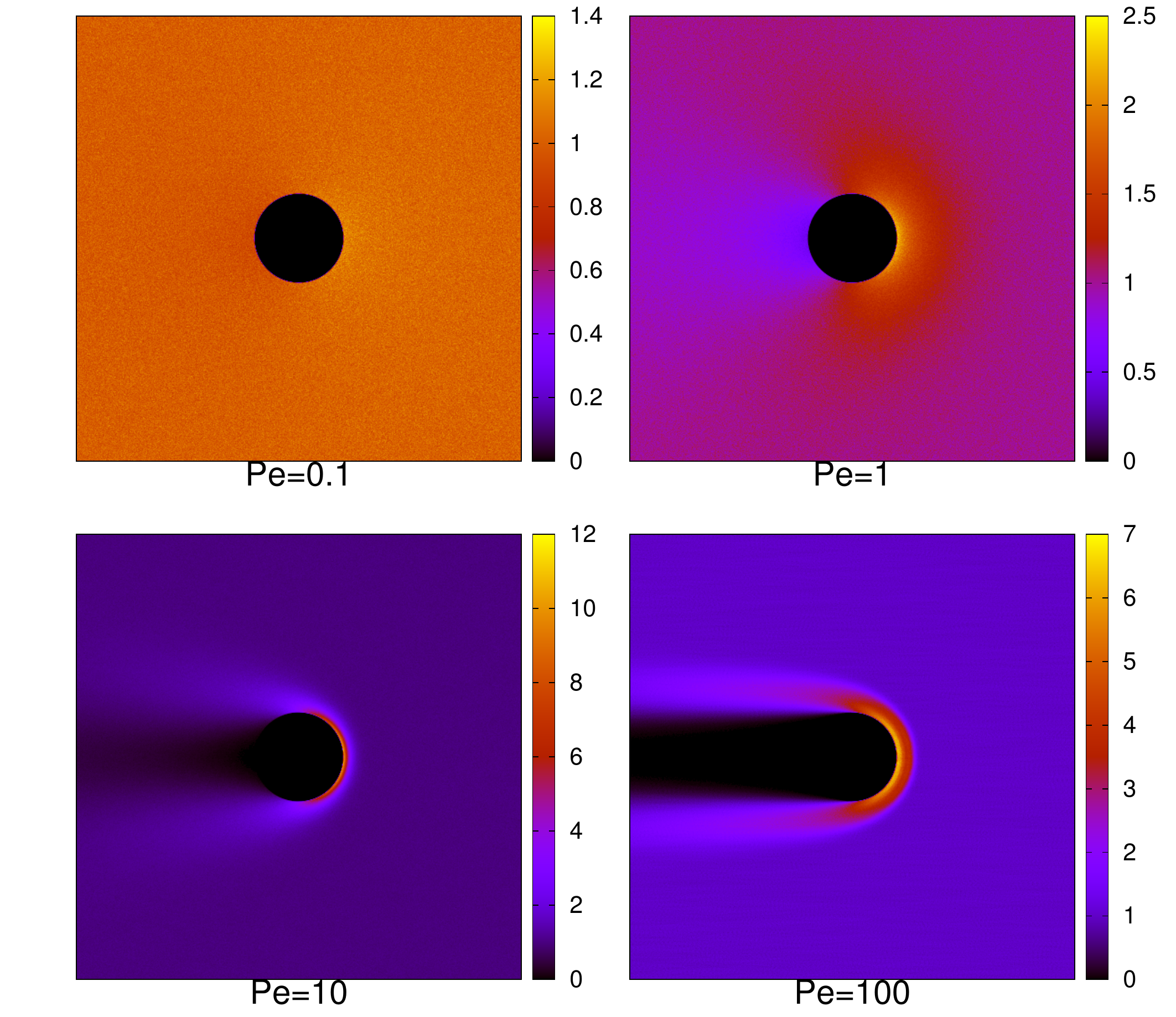} 
\caption{  \label{gr}
Pair distribution function $g(\bm{r})$
 of bath particles in the frame of the probe for different Peclet numbers.  The solvent friction is $\gamma_0=100$ and the temperature is $k_BT=100$. 
}
\end{center}
\end{figure}

\red{
It is convenient to compare \blue{the behaviour of bath particles} in 
different $Pe$ regimes by computing the pair distribution function
$g(\bm{r})$,
which is the normalized number density of bath particles in the coordinate of the probe $g(\bm{r})=n(\bm{r})/n_0=V\int d\bm{v}\, p(\bm{r},\bm{v})$
 ($V$ is the volume of bath particles, here for 2d it is the area, $n_0=N/V$). Fig.~\ref{gr} shows 
 the simulation result $g(\bm{r})$ for different $Pe$ numbers.
For small $Pe$ numbers
$Pe=0.1,1$, bath particles are both built up in front and left behind of the probe, \textit{i.e.} the diffusion dominating regime.} 
As $Pe$ is quite large, $Pe=10,100$, only a thin layer of bath particles build in front but no particles left behind in a long tail region of the probe, which means that the diffusion is ignorable.
\red{
The observation that
 diffusion dominates 
 in the small $Pe$
and is unimportant in large $Pe$,  is consistent with the TTSP.}

\subsection{\blue{velocity distribution}}
\begin{figure}[htbp]
\begin{center}
 \includegraphics[width=0.95\linewidth]{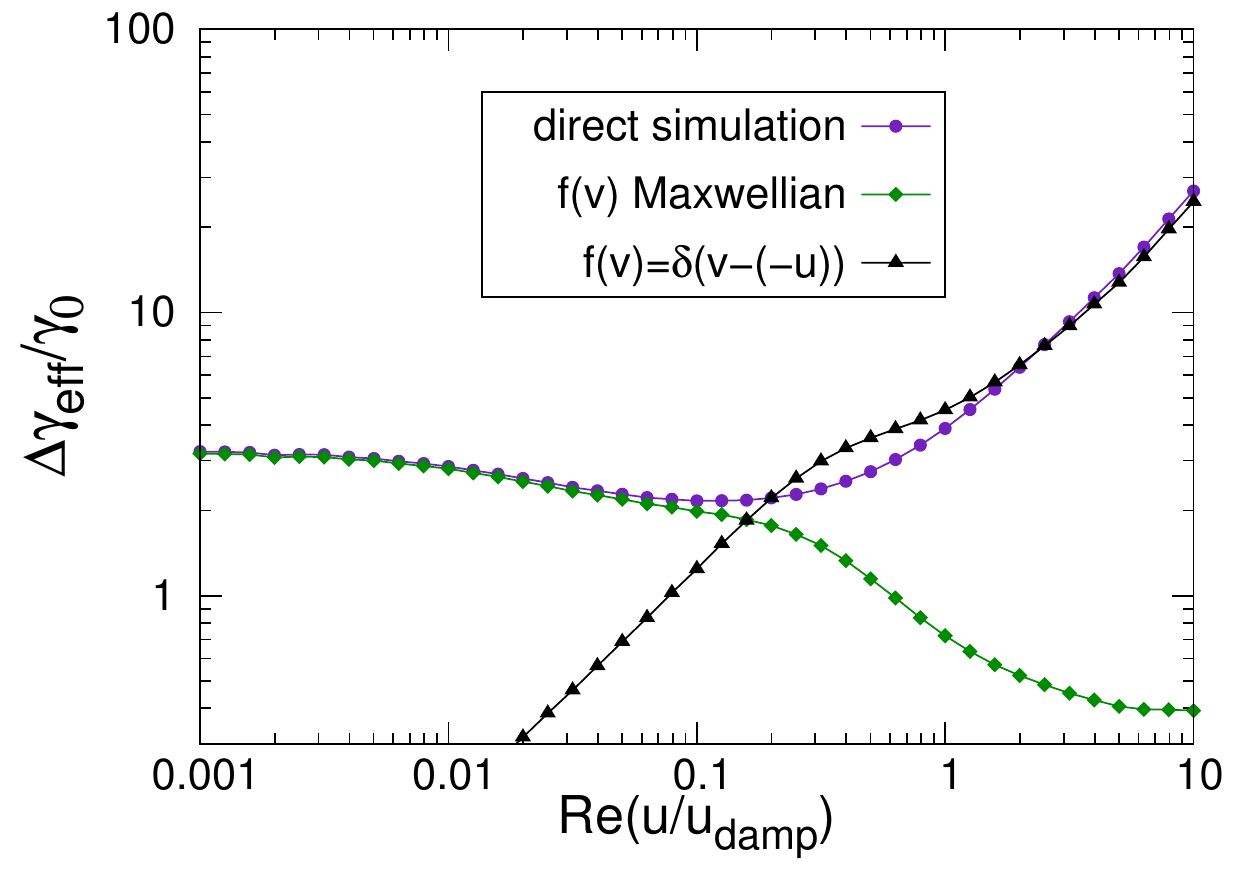} 
\caption{  \label{comp}
Increased effective friction from different methods: the violet line is the direct simulation result.
the green and the black lines are calculated by inputting $g(\bm{r})$ from the simulation associated with Maxwellian distribution and delta distribution of velocity parts, respectively.
}
\end{center}
\end{figure}

Does the pair distribution function contain enough information to calculate the effective friction? If the velocity of bath particles is Maxwellian distributed:
 $f_{eq}(\bm{v})=\big(\frac{1}{\sqrt{2\pi }v_{th}}\big)^d e^{-\frac{\bm{v}\cdot\bm{v}}{2v_{th}^2}}$ with thermal velocity $v_{th}=\sqrt{k_BT/m_b}$,
then the total probability can be separated into
 $p(\bm{r},\bm{v})=V^{-1}g(\bm{r})f_{eq}(\bm{v})$,
and the collision force in Eq. \eqref{fcol}
is reduced to  
\begin{equation}\label{fcol2}
\avg{\bm{F}_{p,col}}=-n_0 k_{B}T\oint_{r=R} dl\bm{e}_n g(\bm{r}),
\end{equation}
which is identical to the one in 
ref. \cite{Squires2005b}.
If the velocity is delta distributed, $f(\bm{v})=\delta
\big(\bm{v}-(-\bm{u})
\big)$, the collision force in Eq. \eqref{fcol}
is reduced to
\begin{equation}
\label{fcol3}
\avg{\bm{F}_{p,col}}=-n_0 2m_b\oint_{r=R} dl\bm{e}_n g(\bm{r})
(\bm{u}\cdot\bm{e}_n)^2
\Theta[\bm{u}\cdot\bm{e}_n]
\end{equation}
Inputting  $g(\bm{r})$
from the simulation into Eq.\eqref{fcol2} and Eq.\eqref{fcol3}, respectively, we obtain  two increased effective frictions, the green line and black line, respectively, as indicated in fig.~\ref{comp}.
Comparing them with the
direct simulation result, the violet line (see
fig.~\ref{comp}),
one can conclude that
\red{ for the calculation of the friction, the pair distribution function still works, but the proper velocity distributions should be input, according to different $Re$ regimes:}
Maxwellian distribution in low Re and the delta distribution
$\delta
\big(\bm{v}-(-\bm{u})
\big)$ in high Re.  


\subsection{
\blue{
streamline
at $T=0$}}
\begin{figure}[t]
\begin{center}
 \includegraphics[width=0.95\linewidth]{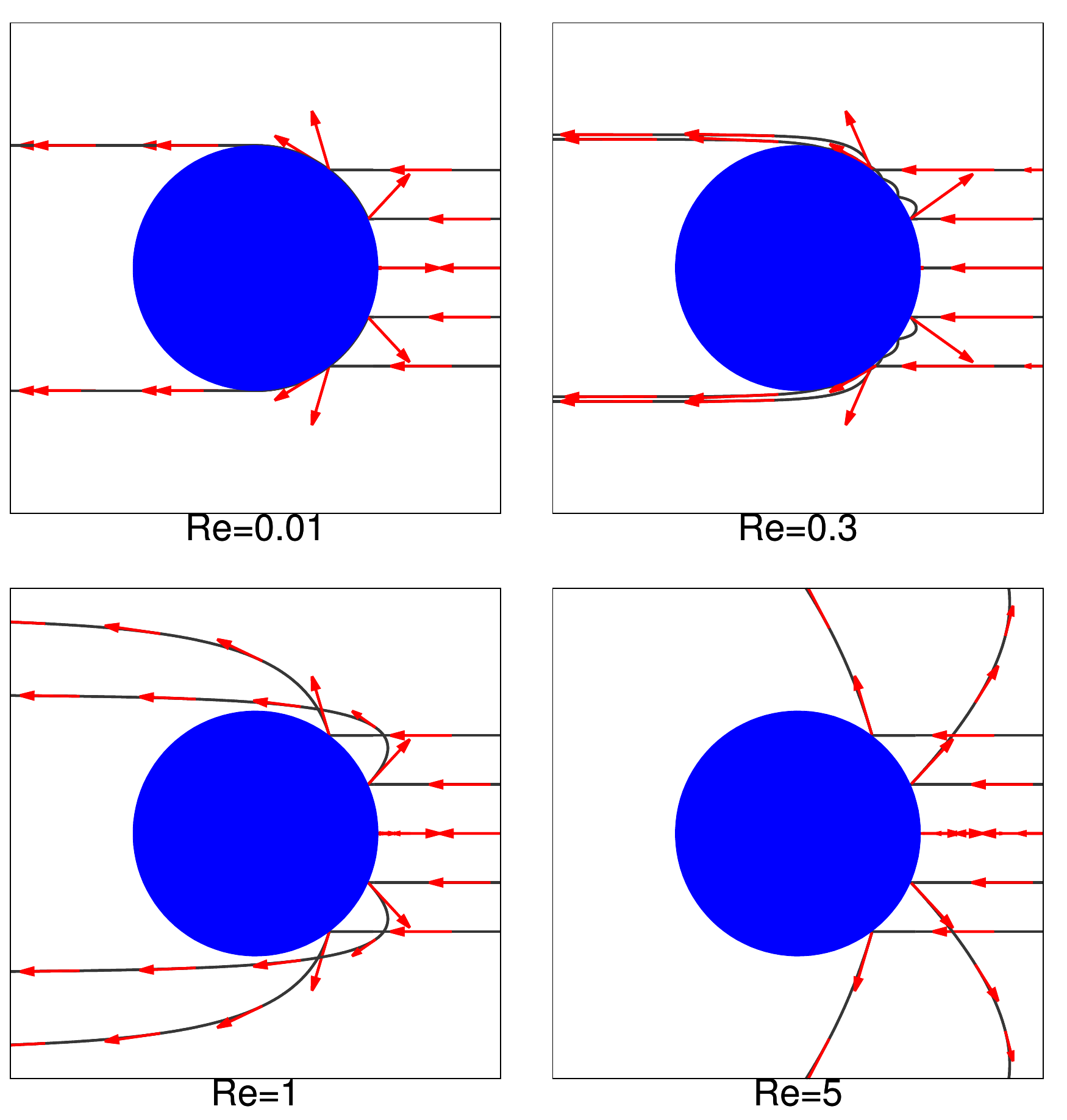}  
\caption{  \label{creep}
Streamlines of bath particles for different Reynolds numbers at the $T=0$ limit. Red arrows are the velocity field (rescaled by the pulling velocity for comparison). For small Reynolds number, the bath particles behave like creep flow around the contact surface. For high Reynolds number, they behave gas-like: single collision and flying away.
The corresponding increased frictions are indicated with
four filled blue circles in
the gray line in  
fig.~\ref{Pe_Re}(c). 
}
\end{center}
\end{figure}

To further investigate the role of the Reynolds number, 
let us consider the $T=0$ limit, 
\blue{where the diffusion time scale is ruled out,}  $Pe\rightarrow\infty$. 
Eq.\eqref{dyb2} is reduced to  
\begin{equation}
 m_b\dot{\bm{v}}=-\gamma_0\bm{v}
 -\gamma_0\bm{u}
\end{equation} 
with the RBC  \eqref{rbc}. 
Interestingly, such simple dynamics
provides a clear mechanism of thickening: the crossover from creep flow in the low Re to gas-like (inertial) flow in high Re, see 
fig.~\ref{creep}. 
The black curves are the streamlines of the
bath particles in the frame of the probe; red arrows are the velocity field.  
Before any collision, bath particles are moving with a constant velocity $-\bm{u}$. Collision causes mirror-like reflection. The term $
-\gamma_0\bm{v}$ reduces the velocity, while  $-\gamma_0 \bm{u}$   
accelerates it. 
 A loose criteria of single-collision-only should be $u t_{damp}\geq R $, \textit{i.e.}  $Re=u \frac{m_b}{R\gamma_0}\geq 1$.
 In the small $Re$ limit, many P-B collisions occur and the bath particles tend to creep along the surface,
see fig.~\ref{creep} $Re=0.01$, 
which causes  $F_{col}\propto  u$ and $\Delta \gamma_{eff}\propto u^0$(the proof will be given somewhere), while in the large Re limit, the single collision causes 
$F_{col}\propto  u^2$ and $\Delta \gamma_{eff}\propto u^1$.  

\blue{Based on the microscopic picture of $Re$ (fig.~\ref{creep}),    
we can also understand why Maxwellian distribution works in low Re but fails in high Re. Let us consider  the $Pe\gg1$ limit, 
where a bath particle moves
with velocity $-\bm{u}$ relative to the probe before any P-B collision
\footnote{
Indeed, before any P-B collision, the motion of a bath particle
is determined by $Pe$ only. It has nothing to do with $Re$. 
}. 
$Re$ determines whether the solvent plays a role during P-B collisions. 
  1) If $Re\ll1$,
damping dominates, the injecting velocity
$-\bm{u}$ of the bath particle is quickly  
 "erased" 
due to the damping and agitation 
processes by the solvent
at the beginning of a few P-B collisions. In the following many times P-B collisions, the bath particle 
transfers the thermalized velocities
to the probe.
 That's why Maxwellian distribution works in this limit.
b) If 
$Re\gg1$, inertia dominates, the P-B collision happens  once only. The bath particle's velocity transferring to the probe  
 is exactly the injecting velocity $-\bm{u}$, which has nothing to do with the solvent. Thus, instead of the Maxwellian, the delta distribution $\delta
 \big(
\bm{v}-(-\bm{u})\big)$ 
works in this limit.
}
  

\section{Conclusion} 
We propose a TTSP to unify thinning and thickening phenomena in active microrheology (see fig.~\ref{1}), and confirm it by a model of pulling with fixed velocity.
The simulation result (fig.~\ref{Pe_Re}), which is equivalent to the solution of the FPE \eqref{fpe0} in steady state, 
 shows linear response,  thinning and thickening.  As far as we know, this is  the first example demonstrating that both thinning and thickening can occur in non-interacting bath particles systems (only P-B interaction is included), which indicates that the many body interaction is not necessary for thinning/thickening behaviour in the low density. 
Furthermore, 
as shown in 
fig.~\ref{Pe_Re},
the
results of the turning points of the thinning and thickening being around $Pe=1$ and $Re=1$, respectively,
and the friction behaviour in different time scale regimes,
 all agree with the TTSP. 

\red{Microscopically,
the pair distribution function $g(\bm{r})$ is  obtained from the simulation as shown in
figs.\ref{gr}.
For the calculation of the friction, we find that with the input $g(\bm{r})$ from the simulation, Maxwellian distribution works in low Re, but fails in high Re; while the delta distribution $\delta\big(\bm{v}-(-\bm{u})\big)$ works in high Re, but fails in low Re.
In the T=0 limit ($Pe\rightarrow\infty$), we obtain a clear microscopic picture of thickening for different $Re$ regimes,  see fig.\ref{creep}. When $Re\ll1$, damping dominates,
the constant friction comes from creep flow, the bath particles collide with the probe and then creep around it; when $Re\gg1$, inertial dominates,
the increasing friction comes from the single P-B collision.
Based on the picture of bath particles in different $Re$ regimes, the validity/invalidity of Maxwellian distribution can also be understood. 
}

\red{ 
According to the TTSP, 
thinning arises from the crossover from diffusion to damping, and thickening arises from the crossover from damping to inertia.
 Note that diffusion was not involved in the experiments of pulling a single particle in static
($T=0$) granular systems
\cite{Takehara2010,
Takehara2014}, that's why thinning was not observed.
  For the same reason, it was
not included in our earlier kinetic model  \cite{Wang2014}.  
Thickening was not found in colloidal systems
\cite{Wilson2009,Gomez-Solano2014,Carpen2005,Squires2005b,
Brader2014}, because 
they were limited to $Re\ll1$ regime, where inertia is unimportant.}

\blue{
The TTSP should also be
valid in the high density with dressed values of $Pe$ and $Re$. B-B many
body interaction increases the    friction of a single bath particle,
$\gamma'_0>\gamma_0$ (in the low density limit, $\gamma'_0$
is just the solvent friction $\gamma_0$). Based on the TTSP, 
the turning point of thinning $Pe=
1\propto
u\gamma'_0$ should shift to a smaller pulling velocity value, 
and 
 that of the thickening
 $Re=
1\propto
u/\gamma'_0$  should shift to a larger value.
}


\acknowledgments
We thank Hailong Peng, Sixue Qin, John Brady,
Thomas Voigtmann, Peidong Yu and 
Hideyuki Mizuno for valuable discussion,
\red{and H. M. and H. P. for critical reading of the manuscript.} 
We acknowledge funding from DFG For 1394 and DAAD.

\bibliography{writing-epl.bib}
\bibliographystyle{eplbib}


\end{document}